# A Price to Enter: Anticipatory Housing Market Sorting and Access Inequality under New York's Congestion Pricing

Mingzhi Xiao[1*], Yuki Takayama[1]

**Abstract:** This study examines how congestion pricing shapes housing market outcomes and spatial equity in New York City. Using high-frequency sales and rental data and a combination of propensity score matching difference-in-differences, geographic regression discontinuity, and event study designs, the analysis identifies distinct short-run adjustment patterns triggered by the policy announcement. Housing prices inside the toll zone fell by about 3.3% and rents by 3%, with the sharpest declines occurring immediately after the announcement. These effects weakened over time, and price resilience emerged among premium properties, indicating early market sorting and growing segmentation. The Geo-RDD results show a clear boundary penalty, with properties just inside the cordon experiencing more pronounced declines than otherwise similar properties just outside. Renters and lower-value segments were more exposed to early adjustment pressures, while implementation-stage effects were limited. The findings suggest that congestion pricing can reshape urban space not only by altering mobility incentives but also by redistributing access and opportunity. Equity-oriented design that includes early-stage support for boundary neighborhoods and renters, along with reinvestment of revenues into untolled transit access, is important for ensuring that the benefits of congestion pricing are shared rather than concentrated.

**Keywords:** congestion pricing; housing markets; announcement effects; boundary effects; spatial equity; geographic regression discontinuity; New York City

# 1 Introduction

Urban congestion pricing has moved from theory to practice as cities seek to manage traffic externalities and improve accessibility. London provides early evidence on measurable gains in traffic conditions and a detailed account of policy rollout and design choices (Leape, 2006; Santos & Shaffer, 2004). Stockholm reports sustained benefits and a mature assessment of costs and outcomes in a setting with strong policy continuity (Börjesson et al., 2012; Eliasson, 2009). Milan adds further implementation experience and highlights the importance of local context and program objectives (Rotaris et al., 2010).

A growing literature studies how transport pricing and related accessibility changes are capitalized into land and housing markets. London and Milan show that improved access and cleaner environments can be reflected in residential prices within charging zones under specific conditions (Tang, 2021; D'Arcangelo & Percoco, 2015). Singapore documents heterogeneous real estate responses that depend on transit baselines and market institutions in both residential and commercial segments (Agarwal et al., 2015). A complementary set of studies uses spatial boundaries and transport innovations to identify localized effects on housing markets. School district and administrative borders reveal sharp price discontinuities that reflect fine scale valuation of neighborhood quality (Black, 1999; Gibbons et al., 2013). New transit options and corridor access shocks further show that boundary placement and connectivity changes translate into local capitalization and spillovers within the urban fabric (Gibbons & Machin, 2005; McMillen & McDonald, 2004; Billings, 2011). Environmental quality shocks also map into housing values and illustrate how amenities and disamenities are priced by households (Chay & Greenstone, 2005; Greenstone & Gallagher, 2008).

[1] Institute of Science Tokyo, 2-12-1 W6-9, Ookayama, Meguro-ku, Tokyo 152-8552, Japan
* Corresponding author: xiao.m.1475@m.isct.ac.jp

Two gaps remain central for understanding distributional consequences in large cities. Most empirical evaluations emphasize post implementation outcomes while housing markets can adjust at the announcement stage when expectations change and uncertainty is high (Agostini & Palmucci, 2008; Chay & Greenstone, 2005). Causal evidence from large and segmented United States housing markets is limited in scope and timing which constrains what we know about anticipatory effects and their incidence across groups and places (Anas & Lindsey, 2011). Equity work shows that who pays and who benefits depends on design choices and on revenue use which calls for analyses that connect pricing details to heterogeneous exposure across the urban space (Eliasson & Mattsson, 2006; Raux & Souche, 2004). Driver behavior and pollution responses under road pricing also suggest that mode dependence can mediate exposure and adjustment paths within cities (Gibson & Carnovale, 2015).

New York City's Central Business District Tolling Program offers a setting that speaks directly to these gaps. The program prices motorized entry into Manhattan's core while public transit remains untolled which creates a mode specific policy environment with a clear geographic boundary (Small et al., 2024). Figure 1 illustrates the Congestion Relief Zone and its boundary, which frames our spatial identification of properties located just inside and just outside the cordon. An extended announcement to activation window allows researchers to separate anticipatory responses from early activation effects using high frequency sales and rental data in a segmented housing market (D'Arcangelo & Percoco, 2015). This institutional structure enables an analysis that centers timing, boundary placement, and access channels in a large and unequal city. This policy, by imposing a direct cost on a previously free movement right, creates a natural experiment to study how the burden of regulation is unevenly capitalized into housing markets and how such capitalization can reinforce preexisting spatial inequalities.

We adopt a simple and transparent empirical design aligned with timing and geography. A geographic regression discontinuity compares properties just inside and just outside the cordon to identify localized price discontinuities near the boundary (Black, 1999). A propensity score difference in differences contrasts treated and closely matched untreated areas to study average effects over time and to support parallel trend diagnostics in a familiar framework (Gibbons et al., 2013). A dynamic event study traces leads and lags around announcement and implementation to separate anticipation from adaptation and to examine heterogeneity by tenure and by price tier in a common timeline design (Agostini & Palmucci, 2008).

Our analysis centers on three questions. We ask whether announcement alone is sufficient to move prices and rents inside the cordon in a setting with mode specific tolling. We ask where and for whom adjustments concentrate with a focus on boundary proximity, tenure, and price tiers. We ask how these patterns relate to access equity in a city where car reliance and transit access vary sharply across neighborhoods, testing the hypothesis that the policy's benefits may accrue to wealthier residents through anticipated congestion relief while the costs are more likely to fall on renters and on households living near the policy boundary.

Finally, the remainder of the paper proceeds as follows. Section 2 reviews research on congestion pricing outcomes, capitalization in housing markets, and equity-oriented design with attention to boundary placement and revenue use. Section 3 describes data sources, variable construction, policy dates, and identification choices for the geographic regression discontinuity, the propensity score difference in differences, and the event study. Section 4 reports baseline estimates and heterogeneous effects across distance bands, tenure, and price tiers and it presents robustness checks on bandwidths, placebo boundaries, and pretrend diagnostics. Section 5 discusses mechanisms that link mode specific pricing to early sorting and access inequality, and it outlines policy implications for announcement phase mitigation, boundary targeting, and revenue recycling. Section 6 concludes with the main findings and directions for future research in large and unequal urban settings.

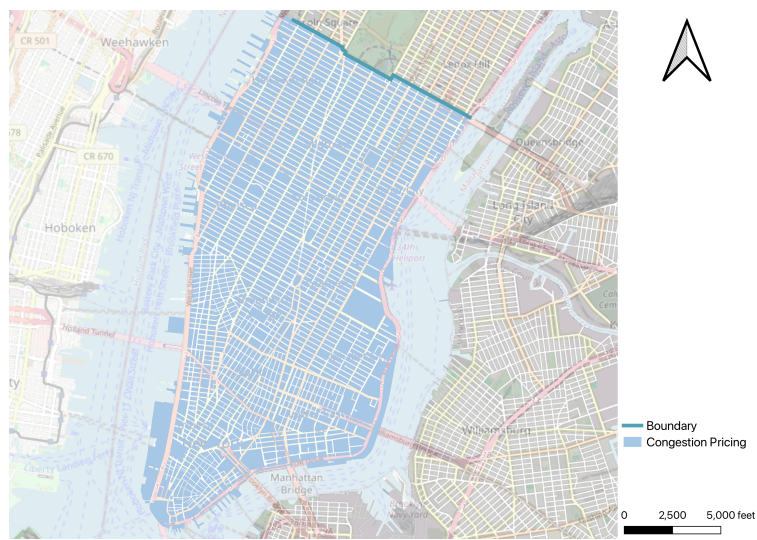

**Figure 1. Congestion Relief Zone within Manhattan's Central Business District**

## 2 Literature Review

Urban congestion pricing has been implemented in many major cities and has generated a sizable evidence base for planners and policymakers. London's central charge reduced traffic volumes and produced implementation lessons that informed later designs (Leape, 2006; Santos & Shaffer, 2004). Stockholm reported sustained reductions in congestion and emissions with public acceptance rising as outcomes became visible (Börjesson et al., 2012; Eliasson, 2009). Milan and Singapore illustrate how impacts vary with policy design and local context. Milan's Ecopass was linked to moderate rent increases within the zone, which is consistent with changes in amenities and access (Rotaris et al., 2010; D'Arcangelo & Percoco, 2015). Singapore's Electronic Road Pricing showed limited changes in residential prices and notable declines in some commercial segments, which underscores the role of transit baselines and market institutions (Agarwal et al., 2015). Taken together, these cases imply that benefits and burdens are not evenly shared, and that equity depends on local design choices and preexisting mobility options.

Beyond transport efficiency and environmental benefits, a growing literature examines capitalization in urban real estate markets. Studies from London and Milan show that improvements in accessibility and air quality can be capitalized into higher residential values under specific conditions (Tang, 2021; D'Arcangelo & Percoco, 2015). Work on spatial boundaries and transport innovations documents pronounced boundary effects and spillovers that can reallocate neighborhood desirability at fine spatial scales (McMillen & McDonald, 2004; Gibbons & Machin, 2005). Anticipation around major transport changes can also move prices before activation, which motivates empirical designs that separate announcement responses from realized effects (Agostini & Palmucci, 2008). These findings suggest that capitalization may reinforce or relax spatial inequality depending on where advantages and costs are localized.

Theoretical research clarifies mechanisms behind these distributional patterns. Classical urban models describe how access and land rents interact in monocentric settings (Alonso, 1964; Mills, 1967; Muth, 1969). More recent contributions incorporate commuter heterogeneity and dynamic adjustment under bottleneck congestion. In a dynamic monocentric model with heterogeneous values of time and flexibility, congestion pricing can shift urban structure in ways that benefit time rich or higher income commuters while raising burdens for others, even when density patterns differ across cities (Takayama, 2020). This framework highlights why mode dependence and income gradients are central to assessing fairness in stratified metro areas.

Despite extensive international evidence, large scale causal analyses in the United States remain limited. Reviews note that much U.S. work relies on simulations or small pilots, which leaves open questions about anticipatory capitalization and heterogeneous incidence in large metropolitan housing markets (Anas & Lindsey, 2011). New York City's program creates an opportunity to study how a mode specific toll with a clear boundary and a long announcement window interacts with local housing markets and neighborhood composition in a complex setting. The policy's mode specificity also allows a direct lens on who bears the driving cost and who can substitute toward untolled transit, which is an equity question at its core.

Methodologically, recent studies employ quasi experimental strategies that are well suited to geographically targeted policies. Geographic regression discontinuity leverages sharp boundaries to identify localized price effects close to the threshold (Black, 1999). Difference in differences compares treated and carefully matched untreated locations over time to recover average causal impacts under parallel trends (Gibbons et al., 2013). Event study designs trace the timing of responses around key policy dates, which helps distinguish anticipation from adaptation and supports heterogeneity analysis by tenure and price tier (Agostini & Palmucci, 2008). In the present context, Geo RDD pinpoints spatial inequity at the policy boundary, DID connects average incidence to comparable neighborhoods, and event study traces differential adjustments between owners and renters as well as between price segments.

Issues of equity and distribution remain central. Who pays and who benefits depends on design and revenue use, with implications for acceptability and welfare incidence across neighborhoods and groups (Eliasson & Mattsson, 2006; Raux & Souche, 2004). Evidence from European programs indicates that revenue recycling toward public transport can offset regressive elements and, in some contexts, yield neutral or progressive net burdens (Börjesson et al., 2012). Studies of behavior and pollution under road pricing highlight the importance of mode dependence and local substitution possibilities for how burdens and benefits are distributed within cities (Gibson & Carnovale, 2015). These insights motivate an evaluation that pays attention to early responses, boundary localization, and segmentation by tenure and value tier.

Based on the above literature, this study focuses on four expectations that guide the empirical analysis in New York City:

1. Announcement effects: Prices within the toll zone may decline around the announcement window as uncertainty rises and access costs are repriced.
2. Heterogeneous impacts: Effects may differ across property segments, with premium units inside the zone showing greater resilience or possible appreciation.
3. Boundary effects: Adjustments are likely to be strongest near the cordon, with local discontinuities and spillovers that weaken with distance.
4. Temporal adjustment: Early declines may moderate as activation approaches and as improvements in accessibility and environmental quality become more evident.

This perspective links anticipated capitalization mechanisms to broader concerns about spatial inequality in a large and segmented housing market.

## 3 Data and Method

To rigorously estimate the causal effect of congestion pricing on housing prices in New York City, we take advantage of a sharp spatial policy boundary and several quasi-experimental methods. Our main empirical strategy is the Geo-RDD, which uses the abrupt spatial change created by the congestion pricing boundary at the edge of Manhattan's Central Business District. This approach makes it possible to credibly identify local treatment effects by comparing housing market outcomes on either side of the policy border. It relies on the assumption that properties just inside and just outside the boundary are otherwise similar (Black, 1999; Small et al., 2007).

Geo-RDD is now widely used in urban economics to assess policy impacts in cases where natural or administrative boundaries produce sudden changes in incentives or amenities (Black, 1999; Gibbons et al., 2013; Small et al., 2007). In urban transportation research, this design has been used to estimate the effects of school quality (Black, 1999), the introduction of new transport infrastructure (Billings, 2011), and environmental policies (Greenstone & Gallagher, 2008) on property values. Our study extends this tradition by applying Geo-RDD to a major policy innovation in the United States.

To strengthen the robustness of our results and address potential threats to identification, we complement the Geo-RDD approach with DID and event study analyses. These methods leverage temporal variation from both the policy announcement and policy implementation. The DID and event study frameworks help control for pre-existing trends, test the parallel trends assumption, and distinguish between anticipatory and realized policy effects (McMillen & McDonald, 2004; Agostini & Palmucci, 2008; D'Arcangelo & Percoco, 2015). By combining these identification strategies, we enhance the credibility of our causal estimates and place our work within the current frontier of policy evaluation research (Imbens & Lemieux, 2008; Small et al., 2007).

In addition, to capture possible differences in treatment effects, we estimate our models separately for neighborhoods with different income levels, distances to the policy boundary, and housing types. This step is particularly important given the international evidence that congestion pricing can have spatially uneven and distributionally heterogeneous impacts (Börjesson et al., 2012; Craik & Balakrishnan, 2022).

## 3.1 Data and Variable Description

Our empirical analysis is based on two complementary data sources. First, we use a panel of individual housing transaction records from the New York City government's open data portal, which provides detailed information on all property sales across the city through December 2024. This dataset includes transaction price, sale date, property attributes, and geographic identifiers, allowing for fine-grained analysis of policy effects at the micro level.

Second, we incorporate the Zillow Home Value Index (ZHVI) and related high-frequency indices published by Zillow Research, available through June 2025. The Zillow Home Value Index is a seasonally adjusted measure of the typical home value in a given geographic area, constructed using the company's proprietary neural network–based Zestimate algorithm. The ZHVI is updated monthly and covers various spatial scales (such as ZIP code and city) and housing types (including all homes, single-family homes, and condos/co-ops). These indices enable us to capture market trends and policy impacts that extend beyond the last available official transaction records.

Descriptive statistics for the main variables by treatment and control groups are presented in Tables 1 and 2. Table 1 summarizes key characteristics during the policy announcement period, while Table 2 provides the corresponding statistics for the policy implementation period. These tables show that, although average housing prices and incomes differ between the toll zone and control areas, the observable characteristics are broadly comparable, supporting the validity of our matched sample for causal analysis.

For the Geo-RDD design, we focus exclusively on properties located within 500 meters of the congestion pricing boundary, using individual transaction data up to December 2024. Table 3 reports descriptive statistics for this subsample, demonstrating the similarity of observable features between properties on either side of the policy boundary and reinforcing the credibility of the regression discontinuity design. To corroborate local identification, we additionally implement a 50-meter donut specification and a placebo Geo-RDD based on pre-announcement transactions; these diagnostics are reported in Appendix A (Tables A.2–A.3). The baseline boundary estimates are summarized in Appendix Table A.1.

For the DID and event study analyses, we employ the full sample of housing transactions from both data sources, comparing treated and untreated areas before and after the policy intervention.

Key explanatory variables include an indicator for location within the congestion pricing zone, time dummies for both the policy announcement and implementation, and a series of property-level controls such as building age, floor area, proximity to subway stations, and school district. All housing price variables are log-transformed to reduce heteroskedasticity and to facilitate the interpretation of coefficients as percentage changes, in line with standard econometric practice (Black, 1999; Chay & Greenstone, 2005). Parallel trends checks are conducted for the DID design, and sensitivity analysis is implemented for the RDD bandwidth selection.

**Table 1. Descriptive Statistics by Group (Policy Announcement)**

| Variable | Control N | Control Mean | Control Std | Control Min | Control Max | Treatment N | Treatment Mean | Treatment Std | Treatment Min | Treatment Max |
|---|---|---|---|---|---|---|---|---|---|---|
| **Housing Price (All Months)** | 3721 | 779,080.23 | 300,583.41 | 194,223.84 | 2,024,646.93 | 550 | 1,543,423.34 | 745,625.16 | 689,355.11 | 3,847,518.29 |
| Area (sqft) | 150 | 4,765,760.49 | 4,567,296.93 | 0.00 | 37,202,637.75 | 22 | 1,053,249.11 | 588,684.20 | 166,789.11 | 2,280,941.93 |
| Health Facilities | 150 | 12.59 | 11.27 | 0.00 | 63.00 | 22 | 11.50 | 8.74 | 0.00 | 27.00 |
| Athletic Facilities | 150 | 43.24 | 33.10 | 0.00 | 143.00 | 22 | 15.09 | 29.40 | 0.00 | 142.00 |
| Childcare Facilities | 150 | 52.87 | 59.87 | 1.00 | 304.00 | 22 | 6.77 | 12.03 | 0.00 | 56.00 |
| Senior Centers | 150 | 1.42 | 1.21 | 0.00 | 5.00 | 22 | 1.23 | 1.63 | 0.00 | 7.00 |
| Hudson Parks | 150 | 0.00 | 0.00 | 0.00 | 0.00 | 22 | 3.50 | 6.79 | 0.00 | 22.00 |
| Libraries | 150 | 1.26 | 0.72 | 0.00 | 3.00 | 22 | 1.09 | 1.02 | 0.00 | 3.00 |
| Food Stores | 150 | 69.07 | 48.67 | 1.00 | 221.00 | 22 | 40.00 | 29.20 | 2.00 | 129.00 |
| Subway Stations | 150 | 3.28 | 3.14 | 0.00 | 16.00 | 22 | 5.27 | 3.25 | 0.00 | 11.00 |
| Public Restrooms | 150 | 6.81 | 3.95 | 0.00 | 18.00 | 22 | 5.55 | 3.98 | 0.00 | 17.00 |
| Bus Stations | 150 | 24.20 | 12.57 | 0.00 | 59.00 | 22 | 16.59 | 12.61 | 1.00 | 40.00 |
| Park Drinking Fountains | 150 | 23.92 | 17.77 | 0.00 | 130.00 | 22 | 17.36 | 19.30 | 0.00 | 93.00 |
| Public Housing Area (sqft) | 150 | 22,735.84 | 71,949.24 | 0.00 | 613,242.14 | 22 | 1,151.31 | 5,404.19 | 0.00 | 24,239.18 |
| Park Area (sqft) | 150 | 1,150,497.38 | 1,653,409.38 | 0.00 | 13,618,478.00 | 22 | 492,107.04 | 710,690.86 | 0.00 | 2,797,062.00 |
| Parking Area (sqft) | 150 | 588,325.32 | 809,509.41 | 0.00 | 7,125,386.00 | 22 | 183,925.77 | 173,011.30 | 0.00 | 599,097.00 |
| Total Population | 150 | 33,663.15 | 30,799.42 | 1,334.00 | 220,510.00 | 22 | 15,119.36 | 8,315.61 | 2,240.00 | 34,899.00 |
| Median Household Income | 150 | 110,227.81 | 52,395.88 | 34,938.00 | 274,424.00 | 22 | 143,049.09 | 41,059.84 | 52,664.00 | 220,592.00 |
| Mean Household Income | 150 | 173,909.91 | 88,922.48 | 59,197.00 | 537,485.00 | 22 | 199,224.27 | 58,968.08 | 90,033.00 | 309,645.00 |
| Households >$100k (%) | 150 | 51.98 | 19.14 | 20.40 | 89.20 | 22 | 65.43 | 12.88 | 34.30 | 82.30 |

**Table 2. Descriptive Statistics by Group (Policy Implementation)**

| Variable | Control N | Control Mean | Control Std | Control Min | Control Max | Treatment N | Treatment Mean | Treatment Std | Treatment Min | Treatment Max |
|---|---|---|---|---|---|---|---|---|---|---|
| **Housing Price (All Months)** | 1950 | 776,500.02 | 288,941.99 | 209,911.82 | 1,901,394.35 | 286 | 1,462,779.37 | 690,100.47 | 692,078.20 | 3,420,498.33 |
| Area (sqft) | 129 | 5,310,443.68 | 4,694,008.77 | 0.00 | 37,202,637.75 | 22 | 1,053,249.11 | 588,684.20 | 166,789.11 | 2,280,941.93 |
| Health Facilities | 129 | 13.13 | 11.63 | 0.00 | 63.00 | 22 | 11.50 | 8.74 | 0.00 | 27.00 |
| Athletic Facilities | 129 | 44.27 | 32.25 | 0.00 | 139.00 | 22 | 15.09 | 29.40 | 0.00 | 142.00 |
| Child Care Facilities | 129 | 55.43 | 62.87 | 2.00 | 304.00 | 22 | 6.77 | 12.03 | 0.00 | 56.00 |
| Senior Centers | 129 | 1.40 | 1.22 | 0.00 | 5.00 | 22 | 1.23 | 1.63 | 0.00 | 7.00 |
| Hudson Parks | 129 | 0.00 | 0.00 | 0.00 | 0.00 | 22 | 3.50 | 6.79 | 0.00 | 22.00 |
| Libraries | 129 | 1.31 | 0.73 | 0.00 | 3.00 | 22 | 1.09 | 1.02 | 0.00 | 3.00 |
| Food Stores | 129 | 72.60 | 50.02 | 2.00 | 221.00 | 22 | 40.00 | 29.20 | 2.00 | 129.00 |
| Subway Stations | 129 | 3.26 | 3.33 | 0.00 | 16.00 | 22 | 5.27 | 3.25 | 0.00 | 11.00 |
| Public Restrooms | 129 | 6.86 | 4.04 | 0.00 | 18.00 | 22 | 5.55 | 3.98 | 0.00 | 17.00 |
| Bus Stations | 129 | 24.34 | 12.88 | 0.00 | 59.00 | 22 | 16.59 | 12.61 | 1.00 | 40.00 |
| Park Drinking Fountains | 129 | 24.07 | 18.03 | 0.00 | 130.00 | 22 | 17.36 | 19.30 | 0.00 | 93.00 |
| Public Housing Area (sqft) | 129 | 22,788.43 | 73,346.18 | 0.00 | 613,242.14 | 22 | 1,151.31 | 5,404.19 | 0.00 | 24,239.18 |
| Park Area (sqft) | 129 | 1,138,144.88 | 1,638,279.23 | 0.00 | 13,618,478.00 | 22 | 492,107.04 | 710,690.86 | 0.00 | 2,797,062.00 |
| Parking Area (sqft) | 129 | 578,271.63 | 795,036.89 | 0.00 | 7,125,386.00 | 22 | 183,925.77 | 173,011.30 | 0.00 | 599,097.00 |
| Total Population | 129 | 35,829.64 | 31,113.48 | 1,334.00 | 220,510.00 | 22 | 15,119.36 | 8,315.61 | 2,240.00 | 34,899.00 |
| Median Household Income | 129 | 116,353.94 | 55,221.38 | 34,938.00 | 274,424.00 | 22 | 143,049.09 | 41,059.84 | 52,664.00 | 220,592.00 |
| Mean Household Income | 129 | 182,010.62 | 93,346.20 | 59,197.00 | 537,485.00 | 22 | 199,224.27 | 58,968.08 | 90,033.00 | 309,645.00 |
| Households >$100k (%) | 129 | 54.36 | 19.22 | 20.40 | 89.20 | 22 | 65.43 | 12.88 | 34.30 | 82.30 |

**Table 3. Descriptive Statistics of Key Variables by Treatment Group (Geo-RDD)**

| Variable | Count (Control) | Mean (Control) | Std. Dev (Control) | Min (Control) | Max (Control) | Count (Treatment) | Mean (Treatment) | Std. Dev (Treatment) | Min (Treatment) | Max (Treatment) |
|---|---|---|---|---|---|---|---|---|---|---|
| SALE_PRICE | 849 | 8,077,113.72 | 59,414,223.09 | 7,668.00 | 931,000,000 | 1655 | 2,783,228.02 | 20,044,914.52 | 12,500.00 | 445,000,000 |
| HubDist | 849 | 341.99 | 151.11 | 24.18 | 556.92 | 1655 | 289.84 | 144.85 | 30.10 | 541.43 |
| SubwaySta_DIS | 849 | 981.86 | 492.96 | 108.20 | 1,992.80 | 1655 | 1289.02 | 618.19 | 120.00 | 2,317.13 |
| PublicRest_DIS | 849 | 953.58 | 470.71 | 161.57 | 2,106.13 | 1655 | 953.17 | 347.69 | 110.96 | 1,610.73 |
| BUS_Sta_Dis | 849 | 677.30 | 438.56 | 84.52 | 1,822.09 | 1655 | 381.75 | 213.62 | 31.20 | 841.69 |
| Park_DIS | 849 | 748.33 | 381.60 | 162.66 | 1,624.93 | 1655 | 626.22 | 390.68 | 76.72 | 1,887.46 |
| FoodRetail_DIS | 849 | 503.73 | 377.17 | 32.85 | 1,741.60 | 1655 | 344.83 | 176.10 | 21.43 | 833.65 |
| ParkingLot_DIS | 849 | 623.02 | 381.00 | 45.85 | 1,519.56 | 1655 | 791.39 | 411.44 | 68.63 | 2,116.56 |
| BicycleParking_DIS | 849 | 181.32 | 99.61 | 43.36 | 508.35 | 1655 | 220.37 | 128.12 | 49.56 | 710.72 |
| YEAR_BUILT | 849 | 1947.71 | 22.64 | 1899.00 | 2009.00 | 1654 | 1950.10 | 20.31 | 1899.00 | 2007.00 |

## 3.2 Propensity Score Matching and Difference-in-Differences (PSM-DID)

To address potential selection bias arising from systematic differences between properties inside and outside the congestion zone, we employ a Propensity Score Matching Difference-in-Differences (PSM-DID) strategy. This two-step approach aims to improve the comparability between treated and control units prior to policy exposure and subsequently estimate the Average Treatment Effect on the Treated (ATT) of the congestion pricing scheme on housing prices.

In the first step, we estimate the propensity score measuring the likelihood of a property being located inside the congestion charging zone based on pre-policy characteristics. Formally, the propensity score for property iii is defined as:

$$e(X_i) = \Pr(Treat_i = 1|X_i) = \Lambda(\alpha + \theta'X_i)$$

where $Treat_i = 1$ if the property lies within the congestion zone and 000 otherwise, $X_i$ denotes a vector of pre-treatment observable characteristics, and $\Lambda(\cdot)$ is the logit link function. The estimation of $e(X_i)$ relies exclusively on pre-policy attributes to avoid incorporating any variables that may themselves be affected by the policy. We impose the common support condition to ensure that comparable treated and control observations are retained. Properties falling outside the overlap of the estimated propensity score distributions are dropped.

Instead of discarding unmatched observations entirely, we implement inverse probability weighting (IPW) to reweight the control group so as to reproduce the distribution of observable characteristics of the treated group. This procedure allows the DID to be estimated on a weighted sample that is balanced in terms of pre-treatment covariates, producing an ATT parameter of interest. The IPW weights are constructed as follows:

$$w_i = \begin{cases} 1, & if\ Treat_i = 1 \\ \frac{e(X_i)}{1 - e(X_i)}, & if\ Treat_i = 0 \end{cases}$$

which assigns larger weights to control observations that more closely resemble treated properties in terms of pre-policy characteristics.

In the second step, we estimate a weighted two-way fixed effects DID regression on the matched (reweighted) sample. The baseline specification takes the following form:

$$\ln(P_{it}) = \alpha + \delta(Treat_i \times Post_t) + \gamma Z_{it} + \lambda_r + \tau_t + \epsilon_{it}$$

where $\ln(P_{it})$ is the logarithm of the transaction price of property $i$ at time $t$, $Post_t$ is an indicator equal to 1 for transactions occurring after the announcement of the congestion pricing scheme, and $Z_{it}$ denotes a vector of time-varying control variables. The term $\lambda_r$ captures unit-specific fixed effects absorbing all time-invariant unobserved heterogeneity across properties, and $\tau_t$ controls for common time shocks affecting all properties. The coefficient of interest, $\delta$, yields the ATT of the policy on housing prices inside the congestion zone. Standard errors are clustered at the property level to account for serial correlation within units over time.

This PSM-DID framework relies on two main identification assumptions. First, conditional on the matched and reweighted sample, the parallel trends assumption requires that in the absence of the policy, treated and control properties would have followed comparable price trajectories. Second, selection into treatment conditional on pre-policy covariates must be unconfounded. By improving covariate balance prior to treatment and then applying DID, this approach mitigates biases arising from non-random assignment of treatment and enhances the credibility of the estimated causal effect of congestion pricing on housing prices.

## 3.3 Geographic Regression Discontinuity Design (Geo-RDD, 500m)

To complement the average treatment effect estimated in the previous section and to capture the localized price shifts at the congestion cordon, we further implement a Geographic Regression Discontinuity Design (Geo-RDD). While the PSM-DID framework identifies the ATT for properties inside the charging zone, the Geo-RDD isolates the price discontinuity at the policy boundary, providing a more localized causal estimate around the threshold where treatment status changes.

Formally, the baseline local linear Geo-RDD specification is defined as:

$$\ln P_i = \mu + \tau D_i + \beta_1 s_i + \beta_2(s_i \times D_i) + \gamma Z_i + \epsilon_i$$

where $D_i$ equals 1 if property iii lies inside the congestion zone and 0 otherwise, and $s_i$ is the running variable measured as the signed distance to the boundary (negative inside the cordon and positive outside). This functional form allows for different slope coefficients on each side of the boundary, ensuring that identification of $\tau$ relies on local variation in treatment assignment rather than functional form assumptions. The term $Z_i$ denotes a vector of property attributes, and $\epsilon_i$ is an error term.

Following best practice in spatial RDD, we estimate local linear models using a triangular kernel that down-weights observations further from the boundary. The main specification uses a symmetric bandwidth of 500 meters on each side of the cordon. We additionally implement robustness checks using alternative bandwidths (200m and 300m), a donut RDD excluding properties within 50m of the boundary to mitigate misclassification, and a placebo boundary test at a pseudo-cordon where no pricing is applied. Standard errors are clustered at the boundary-segment level to account for spatial correlation along the cordon.

Identification in this Geo-RDD relies on the continuity of potential outcomes at the boundary in the absence of the policy. Provided that no other discontinuous policy or structural change coincides with the cordon and that residential sorting is limited at this fine spatial scale, any systematic price jump at the boundary can be attributed to the congestion pricing scheme. By focusing on properties in the immediate vicinity of the cordon, this approach strengthens the causal interpretation of localized price capitalization effects.

### 3.4 Event Study

Finally, to examine the dynamic effects of the policy over time and to validate the critical parallel trends assumption underlying DID, we conduct an event study analysis within the DID framework. This approach involves interacting the treatment indicator with a set of leads and lags relative to the timing of the policy, allowing us to estimate and visualize the treatment effect at each period before and after the event. Formally, the event study model is expressed as:

$$\ln(P_{it}) = + \sum_{k \neq -1} \delta_k (Treat_i \times EventTime_{k,t}) + \gamma X_{it} + \lambda_r + \tau_t + \epsilon_{it}$$

where $\delta_k$ represents the differential treatment effect at time $k$ relative to the policy event. This flexible framework enables the assessment of both pre-existing trends and the temporal evolution of policy effects and is essential for distinguishing true causal impacts from spurious dynamics. The event study thus complements the previous methods and provides additional robustness to the empirical strategy.

Taken together, the combination of these empirical approaches ensures a comprehensive, credible, and nuanced evaluation of the congestion pricing policy's impact on New York City's housing market, addressing both average and local effects as well as temporal dynamics.

## 4 Analysis Result

### 4.1 PSM-DID Results

The PSM-DID analysis, drawing upon Zillow median housing price data, reveals significant variations in housing market reactions around the congestion pricing policy announcement and implementation periods. During the policy announcement phase (Table 4), housing prices within the designated congestion pricing area exhibited a statistically significant decline compared to control areas (DID coefficient = -0.033, p < 0.01). This immediate market response likely reflects residents' anticipation of potential negative impacts, such as increased commuting expenses and reduced accessibility, prompting some households to relocate towards regions outside the policy boundary.

However, the trend changed only marginally upon actual policy implementation (Table 5). Housing prices in the congestion pricing zone stabilized, and the DID results show a very slight but statistically significant increase relative to control areas (DID coefficient = 0.0037, p < 0.01). Although this effect passes the significance threshold, the magnitude is extremely small and likely of limited economic relevance. This suggests that, after the initial uncertainty dissipated and some benefits such as reduced congestion and improved environmental quality became apparent, the market reaction shifted from negative anticipation to a more neutral or mildly positive adjustment. Overall, the data indicate that the realized impact of congestion pricing on local housing prices is minimal, and the policy did not trigger any substantial appreciation within the toll zone.

Moreover, the individual transaction data from NYC government records (Table 6) provide deeper insights by reflecting changes in housing quality and value more precisely. This dataset illustrates a significant positive effect on individual housing transaction prices following policy announcement (DID coefficient = 0.2524, p = 0.017). Such findings imply that higher-value properties within the policy area experienced increased transaction activity or valuation, pointing to improved perceived quality and attractiveness of residences under the new congestion pricing regime.

**Table 4. PSM-DID Regression Results (Policy Announcement-Zillow Data)**

| Variable | Coef. (Std.Err.) | t-value | p-value | 95% CI |
| --- | --- | --- | --- | --- |
| **Treat × Post (DID)** | -0.03303 (0.00646) *** | -5.11 | 0.000 | [-0.04570, -0.02036] |
| Health Facilities | 0.02835 (0.00039) *** | 72.69 | 0.000 | [0.02759, 0.02912] |
| Athletic Facilities | -0.03352 (0.00018) *** | -186.23 | 0.000 | [-0.03387, -0.03317] |
| Childcare Facilities | 0.05228 (0.00033) *** | 159.19 | 0.000 | [0.05164, 0.05292] |
| Senior Centers | -0.14232 (0.00297) *** | -47.91 | 0.000 | [-0.14813, -0.13650] |
| Hudson Parks DIS | -0.02160 (0.00045) *** | -47.87 | 0.000 | [-0.02248, -0.02073] |
| Libraries | 0.10679 (0.00258) *** | 41.39 | 0.000 | [0.10174, 0.11185] |
| Food Stores | 0.03573 (0.00015) *** | 238.20 | 0.000 | [0.03543, 0.03603] |
| Subway Stations | 0.22230 (0.00089) *** | 249.77 | 0.000 | [0.22055, 0.22406] |
| Public Restrooms | 0.10605 (0.00065) *** | 163.15 | 0.000 | [0.10477, 0.10733] |
| Bus Stations | -0.00991 (0.00021) *** | -47.45 | 0.000 | [-0.01033, -0.00949] |
| Park Drinking | -0.04588 (0.00039) *** | -117.90 | 0.000 | [-0.04664, -0.04512] |
| Bicycle Parkings | 0.00102 (0.00001) *** | 102.00 | 0.000 | [0.000999, 0.00105] |
| Public Housing Area sqft | -0.00001 (0.00000) *** | -6.67 | 0.000 | [-0.0000098, -0.0000096] |
| Park Area sqft | 0.00000 (0.00000) *** | 2.80 | 0.005 | [0.0000014, 0.0000014] |
| Parking Area sqft | 0.00000 (0.00000) *** | 4.76 | 0.000 | [0.0000006, 0.0000007] |
| **N (Observations)** | 918 | | | |
| **R²** | 0.990 | | | |
| **Region FE** | YES | | | |
| **Time FE** | YES | | | |

**Table 5. PSM-DID Regression Results (Policy Implementation-Zillow Data)**

| Variable | Coefficient | Std. Error | t-value | P-value | 95% Conf. Interval |
| --- | --- | --- | --- | --- | --- |
| **DID** | 0.003709*** | 0.0006939 | 5.33 | 0.000 | [0.00234, 0.00506] |
| Health Facilities | 0.095697*** | 0.0006741 | 141.96 | 0.000 | [0.0943747, 0.097018] |
| Athletic Facilities | 0.027687*** | 0.0002673 | 103.88 | 0.000 | [0.0273425, 0.0280349] |
| Childcare Facilities | -0.009939*** | 0.0002848 | -34.90 | 0.000 | [-0.0104978, -0.0093808] |
| Senior Centers | -0.143173*** | 0.0023275 | -61.51 | 0.000 | [-0.1477367, -0.1386096] |
| Hudson Parks DIS | 1.486509*** | 0.0113679 | 130.76 | 0.000 | [1.464215, 1.508803] |
| Libraries | -0.534291*** | 0.006688 | -79.89 | 0.000 | [-0.5474082, -0.5211761] |
| Food Stores | 0.018519*** | 0.0002571 | 72.02 | 0.000 | [0.018051, 0.0190236] |
| Subway Stations | -0.064606*** | 0.000958 | -67.05 | 0.000 | [-0.066349, -0.0644645] |
| Public Restrooms | -0.502318*** | 0.005073 | -99.02 | 0.000 | [-0.5122675, -0.4923697] |
| Bus Stations | 0.061593*** | 0.0005036 | 121.24 | 0.000 | [0.0600716, 0.0620417] |
| Park Drinking | 0.175255*** | 0.002437 | 71.93 | 0.000 | [0.1704856, 0.1800253] |
| Bicycle Parkings | -0.016919*** | 0.0001473 | -114.89 | 0.000 | [-0.0172081, -0.0166291] |
| Public Housing Area (sqft) | -0.000297*** | 2.54e-07 | -116.91 | 0.000 | [-0.0000302, -0.0000292] |
| Park Area (sqft) | 0.00665*** | 5.48e-09 | 131.46 | 0.000 | [6.55e-07, 6.75e-07] |
| Total Population | -7.13e-06*** | 3.14e-07 | -22.91 | 0.000 | [-7.75e-06, -6.52e-06] |
| House Holds Median Income | -0.0001178*** | 8.91e-07 | -131.49 | 0.000 | [-0.0001194, -0.0001162] |
| House Holds Mean Income | -0.0001718*** | 1.31e-06 | -131.49 | 0.000 | [-0.0001743, -0.0001693] |
| Income > 100k (%) | 0.3734087*** | 0.0028121 | 132.78 | 0.000 | [0.3678937, 0.3789237] |
| Observations (N) | 2,210 | | | | |
| R-squared | 0.9993 | | | | |
| Adj. R-squared | 0.9993 | | | | |
| Individual FE | YES | | | | |
| Time FE | YES | | | | |

**Table 6. PSM-DID Regression Results (Policy Announcement-NYC Data)**

| Variable | Coefficient | Std. Error | t-value | P-value | 95% Conf. Interval |
|---|---|---|---|---|---|
| DID | 0.2524** | 0.1060 | 2.38 | 0.017 | [0.0446, 0.4602] |
| RESIDENTIAL_UNITS | 0.0026*** | 0.0009 | 3.10 | 0.002 | [0.00097, 0.00432] |
| COMMERCIAL_UNITS | -0.0037** | 0.0017 | -2.11 | 0.035 | [-0.0071, -0.00026] |
| LAND__SQUARE_FEET | -8.5e-6*** | 2.0e-6 | -4.27 | 0.000 | [-0.0000124, -4.6e-6] |
| GROSS__SQUARE_FEET | 1.53e-6*** | 4.7e-7 | 3.27 | 0.001 | [6.1e-7, 2.44e-6] |
| YEAR_BUILT | 0.00695*** | 0.00135 | 5.10 | 0.000 | [0.00478, 0.00963] |
| TAX_CLASS_AT_TIME_OF_SALE | 0.1694*** | 0.0378 | 4.57 | 0.000 | [0.0967, 0.2421] |
| Subway Sta DIS | 0.00053 | 0.00069 | 0.76 | 0.447 | [-0.00083, 0.00189] |
| Public Rest DIS | -0.00035*** | 0.00009 | -3.84 | 0.000 | [-0.00053, -0.00017] |
| BUS Sta Dis | -0.00052 | 0.00127 | -0.41 | 0.681 | [-0.00299, 0.00195] |
| Park DIS | 0.00016* | 0.00009 | 1.71 | 0.087 | [-0.000023, 0.000389] |
| Food Retail DIS | 0.00073*** | 0.00021 | 3.56 | 0.000 | [0.00029, 0.00135] |
| Parking Lot DIS | 0.00035*** | 0.00010 | 3.56 | 0.000 | [0.00016, 0.00054] |
| Bicycle Parking DIS | -0.00885*** | 0.00229 | -3.87 | 0.000 | [-0.0145, -0.00313] |
| Total Population | -7.01e-6*** | 2.0e-6 | -3.43 | 0.001 | [-0.000011, -3.1e-6] |
| House Holds Median Income | 0.0027 | 0.0048 | 0.56 | 0.576 | [-0.00673, 0.00012] |
| House Holds Mean Income | 0.0033 | 0.0133 | 0.25 | 0.802 | [-0.0228, 0.0295] |
| Income > 100k (%) | 0.0216* | 0.0118 | 1.83 | 0.065 | [-0.0138, 0.0447] |
| Observations (N) | 2,882 | | | | |
| R-squared | 0.2431 | | | | |
| Adj. R-squared | 0.2289 | | | | |
| Individual FE | YES | | | | |
| Time FE | YES | | | | |

## 4.2 Geo-RDD: Localized Price Effects at the Policy Boundary (500m)

The Geo-RDD analysis provides a more localized perspective on housing market responses by focusing on price discontinuities at the congestion zone boundary. Prior to the policy announcement, regression results (Table 7, Figure 2) indicate no significant price difference between properties immediately inside and outside the toll zone (treat coefficient = -0.0671, p = 0.217). This finding confirms that, before any public information was released, the housing markets across the boundary were indeed comparable, validating the natural experiment design.

Following the policy announcement, however, a marginally significant negative discontinuity emerges at the boundary (Table 8, Figure 3; treat coefficient = -0.0971, p = 0.066), suggesting that local housing markets adjusted downward in anticipation of the new policy. The market's response was concentrated around the boundary, illustrating heightened sensitivity to expected changes in accessibility and commuting costs. Properties just inside the toll zone faced greater downward price pressure, while those immediately outside potentially benefitted from spillover demand or improved traffic conditions.

To test the robustness of these results, we further estimated a quadratic polynomial Geo-RDD model (Table 9). The findings remain consistent: before the policy, no significant boundary effect is detected, while after the announcement, the negative discontinuity persists (treat coefficient = -0.089, p = 0.070). The close alignment between linear and quadratic model results enhances the credibility of the observed boundary effect and demonstrates the robustness of the spatial discontinuity design. Additional boundary validity checks including a 50-meter donut exclusion and a pre-announcement placebo Geo RDD provide further support that the observed discontinuity is not driven by edge observations or by preexisting price differences across the boundary (Appendix A, Tables A.1 to A.3).

Together, these results highlight the spatial heterogeneity and localized nature of congestion pricing impacts. The sharp policy boundary not only produces direct economic effects within the toll zone but also generates distinct market adjustments in adjacent neighborhoods. These findings provide essential context for understanding the broader market segmentation and distributional consequences analyzed in subsequent sections.

**Table 7. Geo-RDD Regression Results: Before Policy Announcement**

| Variable | Coefficient | Std. Error | t-value | P-value | 95% Conf. Interval | |
|---|---|---|---|---|---|---|
| treat | -0.0671 | 0.054 | -1.238 | 0.217 | -0.173 | 0.038 |
| Hub DIS | -0.00001 | 0.0001 | -0.081 | 0.935 | -0.0002 | 0.0002 |
| Subway Sta DIS | 0.00001 | 0.00005 | 0.180 | 0.857 | -0.0001 | 0.0002 |
| Public Rest DIS | -0.00008 | 0.0001 | -0.828 | 0.408 | -0.0003 | 0.0001 |
| BUS Sta DIS | 0.00001 | 0.0001 | 0.087 | 0.931 | -0.0002 | 0.0002 |
| Park DIS | -0.00002 | 0.0001 | -0.209 | 0.834 | -0.0002 | 0.0002 |
| Food Retail DIS | 0.00005 | 0.00008 | 0.633 | 0.527 | -0.0001 | 0.0002 |
| Parking Lot DIS | -0.00003 | 0.00006 | -0.539 | 0.590 | -0.0002 | 0.0001 |
| Bicycle Parking DIS | -0.00002 | 0.00009 | -0.236 | 0.813 | -0.0002 | 0.0001 |
| YEAR_BUILT | 0.00008 | 0.0004 | 0.224 | 0.823 | -0.0006 | 0.0008 |

**Table 8. Geo-RDD Regression Results: After Policy Announcement**

| Variable | Coefficient | Std. Error | t-value | P-value | 95% Conf. Interval | |
|---|---|---|---|---|---|---|
| treat | -0.0971* | 0.053 | -1.841 | 0.066 | -0.200 | 0.007 |
| Hub DIS | 0.00001 | 0.0001 | 0.085 | 0.932 | -0.0002 | 0.0002 |
| Subway Sta DIS | 0.00001 | 0.00005 | 0.247 | 0.805 | -0.0001 | 0.0002 |
| Public Rest DIS | -0.00008 | 0.0001 | -0.872 | 0.384 | -0.0003 | 0.0001 |
| BUS Sta DIS | 0.00001 | 0.0001 | 0.133 | 0.894 | -0.0002 | 0.0002 |
| Park DIS | -0.00003 | 0.0001 | -0.263 | 0.793 | -0.0002 | 0.0002 |
| Food Retail DIS | 0.00005 | 0.00008 | 0.613 | 0.540 | -0.0001 | 0.0002 |
| Parking Lot DIS | -0.00003 | 0.00006 | -0.483 | 0.629 | -0.0002 | 0.0001 |
| Bicycle Parking DIS | -0.00002 | 0.00009 | -0.181 | 0.856 | -0.0002 | 0.0001 |
| YEAR_BUILT | 0.00009 | 0.0004 | 0.253 | 0.800 | -0.0007 | 0.0008 |

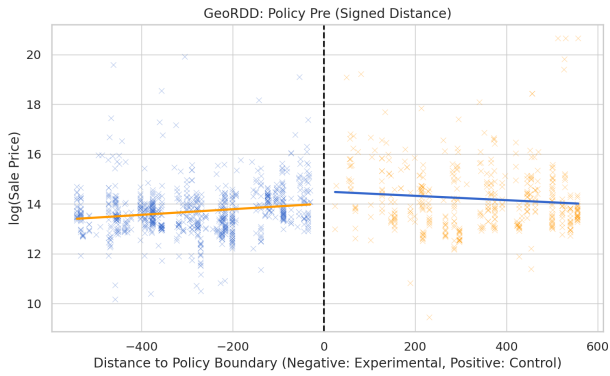

Figure 2. Geo-RDD Before Policy Announcement

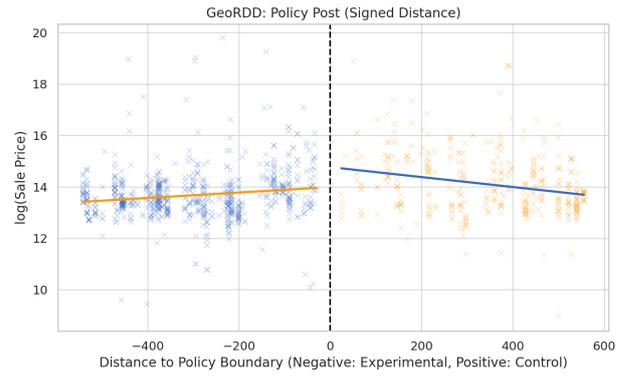

Figure 3. Geo-RDD After Policy Announcement

**Table 9. Quadratic Polynomial Geo-RDD Regression Results**

| | Before Policy Announcement | | | | After Policy Announcement | | | |
|---|---|---|---|---|---|---|---|---|
| **Variable** | **Coefficient** | **Std. Error** | **t** | **P** | **Coefficient** | **Std. Error** | **t** | **P** |
| treat | -0.058 | 0.044 | -1.32 | 0.188 | -0.089* | 0.049 | -1.81 | 0.070 |
| Intercept | 13.437 | 0.193 | 69.75 | <0.001 | 12.750 | 0.186 | 68.51 | <0.001 |
| Hub Dis | 0.0000 | 0.0000 | 0.71 | 0.477 | 0.0000 | 0.0000 | 0.13 | 0.899 |
| Hub Dis_sq | 0.0000 | 0.0000 | 1.12 | 0.263 | 0.0000 | 0.0000 | 0.11 | 0.914 |

## 4.3 Policy Effects on Rental Prices

While the previous sections focused on sales prices, it is equally important to examine the impact of congestion pricing on rental prices, as the rental market often accommodates more mobile and lower-income residents. Table 10 presents the DID regression results for log rental prices, with models estimated separately for the policy announcement and implementation periods. All specifications include individual and time fixed effects to account for unobserved heterogeneity and market-wide shocks. The results show that rental prices within the congestion pricing zone declined by approximately 3% relative to control areas immediately after the policy announcement (DID = -0.0302, p < 0.001). This effect is nearly identical in magnitude to the response observed in the sales market, suggesting that both renters and homeowners reacted similarly to the anticipated impacts of the policy. After policy implementation, the effect on rental prices becomes smaller and is not statistically significant (DID = -0.0053, p = 0.166), indicating that the market quickly adjusted and returned to equilibrium once the initial uncertainty was resolved.

Although the scale of the effect is similar between the rental and sales markets, the rental sector demonstrates a faster and more immediate adjustment. Tenants can change residence or renegotiate leases with fewer barriers than homeowners, allowing the rental market to reflect policy expectations almost in real time. Residents in the rental market also tend to be more sensitive to changes in commuting costs and neighborhood accessibility, making their choices an important indicator of market sentiment. Overall, the findings from the rental market reinforce the evidence from the sales market: congestion pricing primarily leads to a short-term adjustment in housing costs, without causing persistent declines in rents or property values once the policy is in place. These results highlight the importance of timely policy communication and targeted support measures to minimize temporary disruptions, especially for more mobile or vulnerable populations. The next section further explores the mechanisms that drive these housing market responses and considers their broader implications for urban equity and policy design.

In summary, the rental market exhibits an immediate and pronounced response to the announcement of congestion pricing, mirroring the patterns observed in the sales market. However, the faster adjustment in rental prices highlights the heightened vulnerability and mobility of renters, who are typically more sensitive to changes in commuting costs and neighborhood accessibility. This parallel shift across both the rental and sales markets underscores that congestion pricing affects not only asset owners but also day-to-day housing affordability for a broader range of urban residents. Combined with the spatial heterogeneity detected by the Geo-RDD analysis, these findings suggest that rental market adjustments could further intensify boundary effects, potentially increasing residential turnover and short-term rent volatility at the policy edges. Therefore, effective policy communication and timely support for renters are essential to minimize transient disruptions and to promote greater equity as urban markets adapt to new pricing regimes. These implications are further explored in the following sections on mechanisms and policy recommendations.

**Table 10. Effects of Congestion Pricing on Log Rental Prices**

|  | Announce | Implement |
|---|---|---|
| **DID** | -0.0302(0.0037) | -0.0053(0.0039) |
|  | *** |  |
| **t** | -8.27 | -1.39 |
| **P>|t|** | 0.000 | 0.166 |
| **[95% Conf. Interval]** | [-0.0374, -0.0230] | [-0.0129, 0.0022] |
| **N** | 2904 | 1446 |
| **R²** | 0.9937 | 0.991 |
| **Region FE** | YES | YES |
| **Time FE** | YES | YES |

## 4.4 Event Study

The event study analysis provides a nuanced temporal perspective on housing market responses to congestion pricing. Initial parallel trend checks (Figures 4 and 5) confirm that treatment and control groups exhibited similar housing price trajectories before policy events, supporting the DID methodological assumptions. The dynamic event study results (Figures 6 and 7) reveal an immediate negative market reaction following the policy announcement, aligning with residents' anticipations of increased costs and decreased accessibility.

Notably, the event study demonstrates a significant recovery and gradual improvement in housing prices post-implementation, suggesting market resilience and adaptability. The Zillow event study data (Figure 6) highlights how the initial negative reaction peaked shortly after the announcement, then gradually reversed as policy implementation approached and tangible benefits became clearer. Similarly, NYC individual transaction data (Figure 7) reinforces this pattern, illustrating that initial negative expectations transitioned towards positive recognition of policy-driven improvements in local amenities and reduced congestion.

Collectively, these findings underscore the importance of distinguishing between market anticipation and realization effects. The initial market pessimism evident during the announcement phase contrasts sharply with the subsequent positive adjustments post-implementation. This analysis thus provides comprehensive evidence highlighting both short-term anticipatory reactions and long-term adaptations, revealing the complexity of urban housing market responses to significant policy interventions such as congestion pricing.

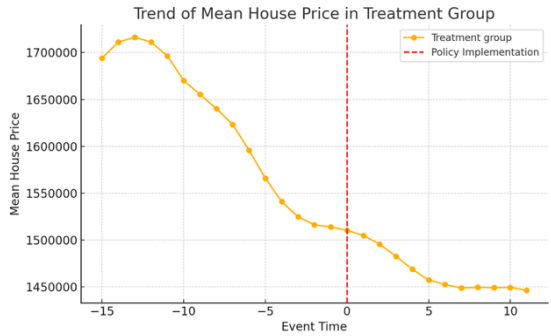

**Figure 4. Mean House Price Trend in Treatment Group**

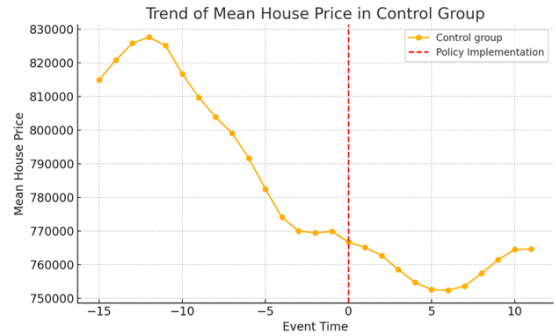

**Figure 5. Mean House Price Trend in Control Group**

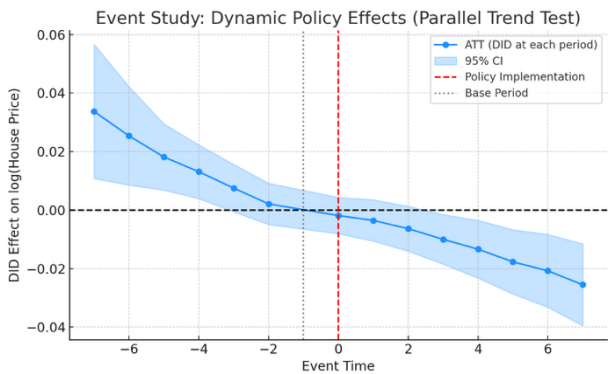

**Figure 6. Event Study (Zillow data)**

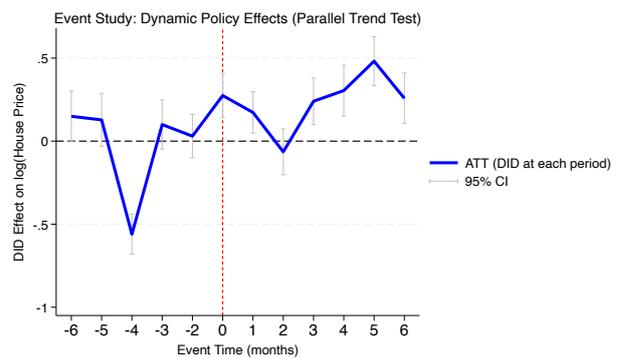

**Figure 7. Event Study (NYC data)**

# 5 Discussion

We find that housing markets respond primarily at the announcement stage and that these responses are concentrated near the boundary and among renter and mid to lower value segments. Expectations, rather than realized charges, dominate short run capitalization in this setting.

Event study estimates in Figure 6 and Figure 7 display a clear break after the announcement and a gradual moderation as the activation date approached. The pattern is consistent with markets capitalizing expected access costs and uncertainty before charges begin, in line with earlier evidence on timing of capitalization (Chay and Greenstone, 2005; Agostini and Palmucci, 2008). These dynamics show that credible policy signals, rather than formal implementation, trigger market adjustments.

A notable feature of the results is the difference between the representative index and the micro transactions. The index for properties inside the cordon declines around the announcement window, while administrative micro data show an increase driven by a shift toward upper tier trades. The index smooths the distribution and adjusts with lags. The micro data record who trades in short windows and therefore reveal sequencing and composition. Quality adjusted comparisons and repeat sales style checks reduce this gap and support the view that composition and timing account for most differences. Figure 6 can illustrate this contrast by placing the index and a quality adjusted measure side by side.

Boundary focused estimates further clarify the mechanism. The geographic regression discontinuity shows a negative discontinuity just inside the cordon that is confined to a narrow spatial band. Thin administrative lines can function as meaningful economic borders when access is repriced locally. The estimates are stable across reasonable bandwidths and functional forms, and placebo boundaries outside the zone do not display comparable breaks. A bandwidth plot such as Figure 8 can highlight the localized nature of the effect and the robustness of the estimates.

Differences between renters and owners add a distributional dimension. Rents inside the cordon fall at the announcement stage and show limited movement at activation, while prices adjust more slowly. This pattern is consistent with faster pass through and tighter budget constraints in rental markets. It also aligns with a mode specific policy design that charges motorized entry while public transit remains untolled, which makes car dependent renters near the boundary more exposed to early repricing of access. Figure 7 can highlight the earlier and sharper rental adjustment relative to sales.

These findings correspond with models that incorporate heterogeneous values of time and commuting flexibility. Agents with higher time values and better substitution options adjust earlier and absorb smaller losses. More constrained groups adjust later and bear larger transitional burdens. The observed sequence of early premium entry, narrow boundary penalties, and stronger rental responses therefore links access cost shocks to short run spatial stratification in a way that theory predicts (Takayama, 2020).

The patterns above carry important implications for fairness and policy design. The announcement window emerges as a critical period for distributional exposure because markets respond once information becomes credible. The boundary belt also deserves attention because localized repricing of access can shift neighborhood desirability over very short distances. Without targeted mitigation, these forces risk reinforcing existing spatial and socioeconomic inequalities. Effective measures include temporary support during the announcement period for boundary residents, visible and near-term transit and last mile improvements that reduce car dependence at the edge, and transparent use of revenues for communities with fewer transport substitutes. Such approaches can maintain the efficiency goals of congestion pricing while reducing unequal burdens.

There remain several areas where further analysis can strengthen the evidence.
1. Composition and level effects could be separated more clearly through quality adjusted indices, quantile models, and repeat sales measures reported alongside the representative index.
2. Boundary sensitivity could be assessed through bandwidth scans, alternative polynomial orders, and placebo boundaries, with and without local covariates.
3. Parallel trends checks should be shown by tenure and value tier, and matching balance should be reported for the difference in differences sample.
4. Alternative timing definitions around the announcement could account for information diffusion and data reporting cycles.
5. If feasible, a proxy for car dependence at a fine spatial scale would help test whether mode based exposure explains the stronger rental and boundary responses.

Taken together, the evidence supports a consistent interpretation. Congestion pricing reshapes housing markets first through expectations and only later through realized charges. Effects are sharp at boundaries, stronger for renters, and heterogeneous across value tiers. The difference between index and micro measures reflects sequencing and composition rather than measurement error. These patterns link theoretical predictions of heterogeneous commuter responses to observed spatial stratification and point toward policy designs that address both timing and location of exposure.

# 6 Policy Implications

Figure 8 provides a clear mechanism for how congestion pricing affects housing markets. It shows that changes in access costs trigger different behavioral responses across groups, which then translate into uneven price and rent adjustments. This link between mechanism and outcome offers direct guidance for policy design.

The first implication concerns when to intervene. Our event study results show that market responses begin at the announcement stage, not at activation. If support arrives only after the charge starts, it will miss the moment when expectations form and are priced into housing outcomes. Early and temporary assistance during the announcement period would therefore be more effective. This could include commuter credits for car dependent workers and short-term rent support for renters living inside the cordon. International experience indicates that policies with early support tend to have smoother adjustment and higher acceptance, as observed in London and Stockholm (Leape, 2006; Börjesson et al., 2012). Acting early signals fairness and can prevent initial shocks from becoming long lasting inequalities.

A second implication concerns where support is needed most. Our Geo-RDD results show that the sharpest effects are concentrated just inside the boundary, meaning that a thin policy line becomes a meaningful economic divide. Targeting support to this narrow zone is therefore more efficient than spreading assistance broadly. Measures could include highly visible transit and last mile improvements for neighborhoods just inside the cordon, a short phase in period for these areas, or distance-based rebates that reduce exposure for households closest to the boundary. Evidence from Milan highlights that attention to these small areas can help avoid concentrated displacement pressures (D'Arcangelo and Percoco, 2015).

A third implication concerns who should receive priority support. Our DID findings show that renters and households in lower value properties experience the earliest and strongest impacts. These groups often have fewer alternatives, weaker financial buffers, and higher sensitivity to travel costs. Targeted measures can protect these residents more effectively than uniform support. Beneficiaries should include renters with limited income who rely on car access for employment, as well as lower value property owners whose equity is most vulnerable to early price adjustments. Tailored assistance can include means tested commuter credits, short term rent support tied to the announcement period, and improved untolled public transport options in renter dense blocks. Clear and visible use of revenues for these purposes strengthens both credibility and fairness, consistent with lessons from earlier cases (Eliasson, 2009).

Sustained monitoring and adaptive adjustments can help maintain equity as the system evolves. Tracking rents, vacancies, and travel patterns in boundary neighborhoods will allow for timely responses if pressure increases. Preannounced rules for adjusting support can improve transparency and build trust. Coordinating with land use policies, such as allowing modest infill or flexible space for services near the boundary, can also help absorb pressure without excluding existing residents.

Taken together, the findings point to a simple but powerful approach. Supporting households early, focusing spatially on boundary neighborhoods, and prioritizing renters and lower income households offers a practical path to a fairer congestion pricing system. If designed with these principles, congestion pricing can improve traffic conditions and environmental quality while still protecting the right to equitable access to the urban core. In this way, efficiency and fairness can reinforce rather than compete.

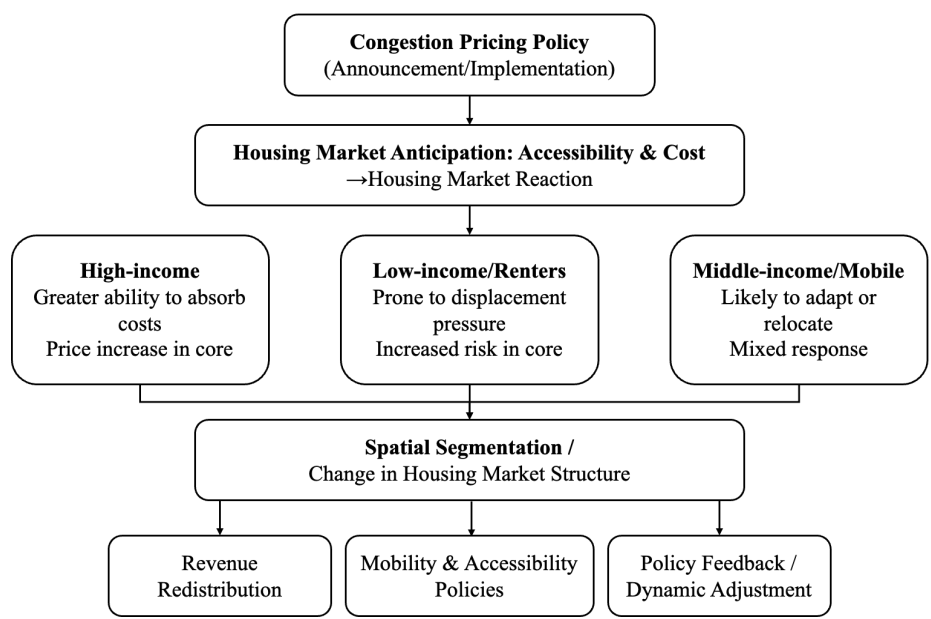

**Figure 8. Mechanism Linking Congestion Pricing Policy to Housing Market Segmentation and Social Equity**

# 7 Conclusion

This paper shifts the focus from the well documented transport benefits of congestion pricing to its social and distributional consequences within cities. Figure 9 summarizes the core findings and makes three patterns visible at a glance: market adjustments begin at the announcement, effects are concentrated at the boundary, and impacts are segmented by tenure and value tier. Prior work shows that pricing reduces congestion and environmental pressures (Börjesson et al., 2012; Leape, 2006). Our results add how these gains are transmitted through housing markets in uneven ways across space and groups.

We find that prices and rents inside the cordon move at the announcement rather than at activation. The decline in representative indices is consistent with higher expected access costs and uncertainty, while micro transactions reveal early entry by premium segments that partially offsets declines through composition. This combination points to expectation driven capitalization and to sequencing in who adjusts first. The pattern implies that policy timing matters for distributional exposure.

Distributional differences are central. Rental markets respond earlier and more strongly than owner markets, which is consistent with greater mobility, tighter budgets, and faster pass through for renters. Within the ownership market, premium properties show more resilience than lower value properties. Together these results indicate that early burdens fall more on renters and on households in lower value segments, raising concerns about who bears the first wave of adjustment (Craik & Balakrishnan, 2023).

Space also matters. The Geo RDD results show a localized penalty just inside the boundary, which means that a thin administrative line operates as a meaningful economic divide. Residents on the edge of the cordon face different housing market conditions from otherwise similar neighbors just outside. Such discontinuities are consistent with prior evidence on boundary capitalization and point to the need for spatially targeted responses (D'Arcangelo & Percoco, 2015; Rotaris et al., 2010).

These findings support a two-part view of congestion pricing. It remains an effective tool for managing traffic and improving environmental quality. It can also reallocate urban advantages if design does not account for timing, location, and incidence. Equity improves when revenues are visibly reinvested in untolled access, when support is offered during the announcement window, and when boundary neighborhoods and renters receive priority in transitional measures (Eliasson & Mattsson, 2006).

The broader implication is straightforward. Cities can capture the efficiency benefits of congestion pricing while safeguarding inclusive access if policy design follows the evidence: act early at announcement, focus spatially on boundary belts, and prioritize renters and lower income households in support and revenue use. With transparent communication and ongoing monitoring, congestion pricing can deliver cleaner mobility and a fairer urban core rather than a more exclusive one.

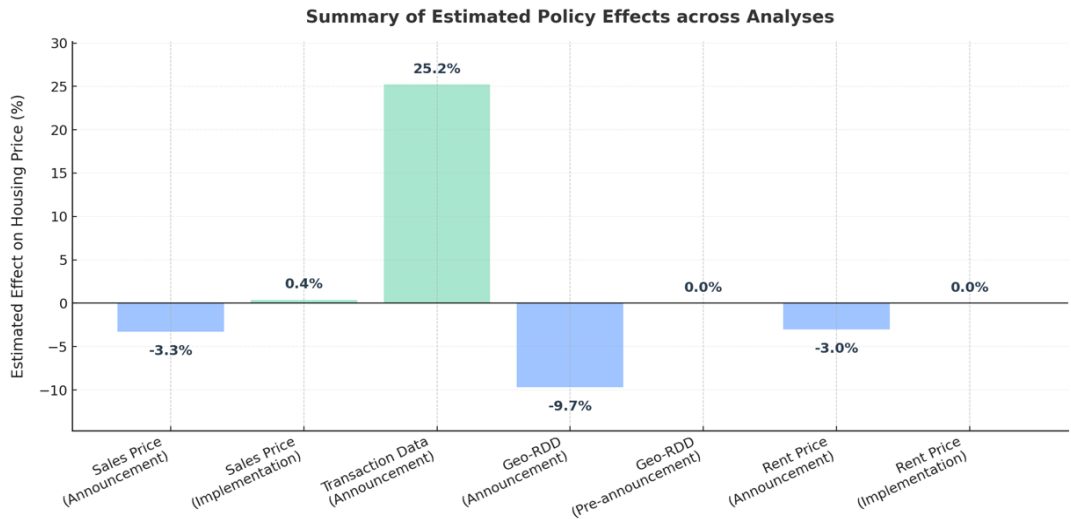

**Figure 9.** Summary of estimated policy effects across analyses.

# 8 Limitations

This study provides evidence on how congestion pricing shapes housing markets and spatial equity in New York City, yet several limitations should be considered when interpreting the results. The analysis centers on housing prices, rents, and spatial distribution, rather than on traffic flows, travel behavior, or environmental outcomes. Earlier research has shown that congestion pricing can reduce traffic and improve air quality (Leape, 2006; Börjesson et al., 2012), so the findings here should be viewed as a complement to transportation focused evaluations. Understanding both mobility and distributional channels is necessary for a full assessment of the policy.

Methodologically, the study relies on the sharp policy boundary and clear event timing used in the Geo RDD and DID designs. These approaches help isolate localized and announcement driven effects, but unobserved neighborhood changes or concurrent policies may still influence the results. In addition, the analysis captures outcomes visible through recorded market transactions and rental listings. It may not detect off market adjustments, informal rental agreements, or shifts in preferences and mobility that occur without a transaction (Ahlfeldt et al., 2015).

The results reflect short to medium term responses because the observation period is limited to the periods around the announcement and activation. Long term adaptation, neighborhood change, and demographic adjustment may unfold gradually and are not captured in the current analysis. Future work should extend the time horizon to assess whether early spatial and social effects persist or attenuate.

Although this study improves on earlier work by examining both ownership and rental markets, it cannot fully disentangle impacts on specific socioeconomic groups. Data limitations prevent direct linkage between transactions and household characteristics such as income, tenure history, or vulnerability to displacement. The findings suggest that renters and lower value segments face greater exposure, but household level evidence is needed to confirm how low income residents or long term tenants experience the policy (Craik and Balakrishnan, 2023).

Finally, the findings reflect the context of New York City, where high density, extensive public transit, and a segmented housing market shape responses to congestion pricing. These features may limit the degree to which the results apply to other cities. Applying this framework to different institutional settings and pricing designs would help assess the broader relevance of the mechanisms identified here.

Future research could address these limitations by integrating transportation and environmental data, extending the temporal scope, and linking household level information to housing market outcomes. Such efforts are important for developing a more complete understanding of how congestion pricing affects urban equity and for informing policy designs that balance efficiency with fair access.

# Appendix

This appendix provides a single summary of supplementary boundary diagnostics and estimates that complement the main text. We report baseline Geo RDD estimates at ±300 meters and ±200 meters using a local linear specification with side specific slopes and heteroskedasticity robust standard errors, with log price as the dependent variable. Table A.1 shows a sizable negative discontinuity inside the cordon, with τ around −0.76 at ±300 meters and −0.83 at ±200 meters, which corresponds to an approximate price change of −53 to −56 percent. To assess identification, we implement two compact checks. First, a donut exclusion that removes observations within 50 meters of the boundary yields estimates that remain negative and economically large, as reported in Table A.2, which indicates that transactions immediately at the edge do not drive the result. Second, a placebo Geo RDD that uses only pre-announcement transactions before July 2023 produces small and imprecise jumps, as reported in Table A.3, which indicates that a boundary gap is not present prior to the policy timeline. Together these results support that the discontinuity emerges with policy information rather than from pre-existing structural differences across the boundary and they are consistent with the broader timing and distributional patterns documented in the paper.

**Table A1. Geo-RDD Estimates at the Cordon Boundary (±300m & ±200m)**

| Bandwidth (m) | N | τ (log price) | Std. Error | p-value | $R^2$ | %Δ Price |
|---|---|---|---|---|---|---|
| ±300 | 1,232 | -0.755*** | 0.222 | 0.0007 | 0.137 | -53.0% |
| ±200 | 593 | -0.831** | 0.370 | 0.0247 | 0.046 | -56.4% |

**Notes:** Local linear Geo-RDD with side-specific slopes and HC1 robust standard errors. "%Δ Price" is calculated as $(e^\tau - 1) \times 100$. *** p<0.01, ** p<0.05, * p<0.10.

**Table A2. Robustness Check — Donut Exclusion (|distance| < 50m removed)**

| Bandwidth (m) | N | τ (log price) | Std. Error | p-value | $R^2$ | %Δ Price |
|---|---|---|---|---|---|---|
| ±300 (donut 50m) | 1,115 | -0.701** | 0.301 | 0.0203 | 0.115 | -50.4% |
| ±200 (donut 50m) | 514 | -0.792* | 0.421 | 0.0628 | 0.038 | -54.5% |

**Notes:** Observations within 50 meters of the boundary are excluded. Estimates remain negative and sizable, confirming that the discontinuity is not driven by transactions extremely close to the boundary.

**Table A3. Robustness Check — Placebo Using Pre-Announcement Sales Only (< July 2023)**

| Bandwidth (m) | N | τ (log price) | Std. Error | p-value | $R^2$ | %Δ Price |
|---|---|---|---|---|---|---|
| ±300 | 1,022 | -0.192 | 0.258 | 0.4612 | 0.012 | -17.5% |
| ±200 | 501 | -0.245 | 0.376 | 0.5144 | 0.008 | -21.7% |

**Notes:** Pre-announcement Geo-RDD estimated using sales before July 2023. No significant discontinuity is found, consistent with the hypothesis that boundary effects materialized following the policy announcement.